# Spectropolarimetry of Wolf–Rayet Stars in the Magellanic Clouds: Constraining the Progenitors of Gamma-ray Bursts

Jorick Vink[1]

[1] Armagh Observatory, College Hill, Northern Ireland, United Kingdom

Wolf–Rayet stars have been identified as objects in their final phase of massive star evolution. It has been suggested that Wolf–Rayet stars are the progenitors of long-duration gamma-ray bursts in low metallicity environments. However, this deduction has yet to be proven. Here we report on our initial results from a VLT/FORS linear spectropolarimetry survey of Wolf–Rayet stars in the Magellanic Clouds, which is intended to constrain the physical criteria — such as weaker stellar winds, rapid rotation, and associated asymmetry — of the collapsar model. Finally, we provide an outlook for polarisation studies with an extremely large telescope.

Stars like the Sun spin slowly, with speeds of only a few kms$^{-1}$. By contrast, massive stars rotate much more rapidly, reaching rotational velocities of over 400 kms$^{-1}$. This rapid rotation is understood to have dramatic consequences for their evolution and ultimate demise, which may involve the production of a long-duration gamma-ray burst (long GRB), the most intense type of cosmic explosion since the Big Bang.

## Rotating massive stars

The evolution of massive stars is thought to be the result of a complex interplay between mass loss, rotation and possibly magnetic fields. Whilst the importance of mass loss was established in the 1970s, following the discovery of mass loss from normal O-type supergiants and radiation-driven wind theory, the role of rotation was only fully appreciated in the 1990s (for example, Langer, 1998; Maeder & Meynet, 2000).

When a star rotates, the pole becomes hotter than the stellar equator (Von Zeipel theorem), which enables a rather complex meridional circulation in the stellar interior (see Figure 1). In this process, nu-clear processed material is transported from the core to the stellar envelope, thereby enriching the stellar surface with elements such as nitrogen, which is produced during the CNO cycle of hydrogen burning (in contrast to the proton–proton cycle that operates in solar-type stars).

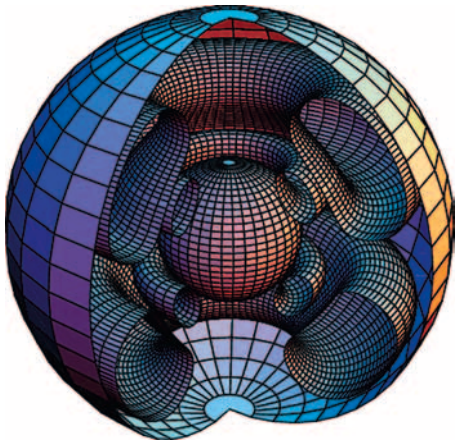

Figure 1. Model of meridional circulation in a stellar interior (from Maeder & Meynet, 2000).

During later evolutionary phases, the combination of mass loss and rotation also leads to the transfer of products of helium burning to the surface, enriching the atmosphere with carbon during the final carbon-rich Wolf–Rayet (WC) phases before the stellar core is expected to collapse, producing a supernova (SN) — in some cases in conjunction with a long GRB.

## The collapsar model for gamma-ray bursts

From the 1960s onwards, GRBs were discovered appearing from all cosmic directions. However, an explanation for their origin was still to be found. A massive breakthrough occurred in 1998 when a European team led by graduate student Titus Galama discovered that the unusual supernova 1998bw fell within the error box of GRB980425, which was subsequently confirmed with the case of SN2003dh/GRB030329. This was convincing evidence that long (those lasting longer than two seconds) GRBs were associated with the deaths of massive stars.

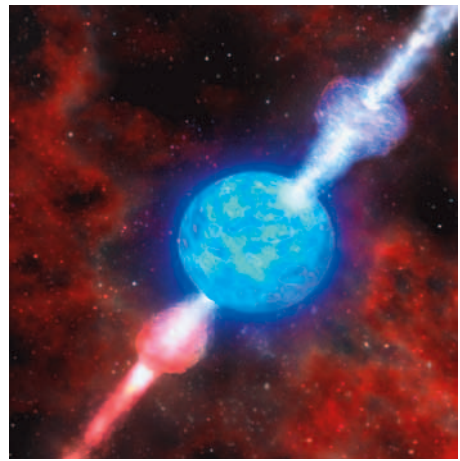

Figure 2. Artist's impression of a gamma ray burst.

The most popular explanation for the long GRB phenomenon is that of the collapsar model (Woosley, 1993), where a rapidly rotating massive core collapses, forming an accretion disc around a black hole. In this process, part of the gaseous material is ejected in the form of two relativistic jets, which are aligned with the rotation axis of the dying star (see Figure 2). These jets are thought to involve an opening angle of just a few degrees, and only when these jets happen to be directed towards Earth are we able to detect the event as a gamma-ray burst.

## The gamma-ray burst puzzle

One of the persistent problems with the collapsar model was that it not only required the star to have a high rotation speed initially, but that the star needs to maintain this rapid rotation until the very end of its life. The reason this is such a challenge is that one of the most characteristic features of massive stars, their strong stellar outflows, are expected to remove angular momentum. Most stellar evolution models show that as a result of mass loss, the objects not only remove up to 90 % of their initial mass when reaching their final Wolf–Rayet (WR) phase, but as a result of this wind, the stars are expected to come to an almost complete standstill. This property seems to be supported by the fact that most Galactic WR winds are found to be spherically symmetric (Harries et al., 1998). Because of the observed spheric-ity, which is thought to be characteristic for their slow rotation, we would not anticipate Galactic WR stars to produce a GRB when they expire.

The question is what do we expect for WR stars in low-metallicity galaxies? Do WR stars in low-metallicity galaxies suffer from similar mass and angular momentum loss? To address these issues we need to explore the underlying physical origin of massive star winds.

Iron iron iron

Radiation hydrodynamic simulations show that stellar winds from massive O-type stars are driven by the radiation pressure on metal lines, and specifically on iron (Fe), despite the fact that it is such a rare element. In the Milky Way, for each and every Fe atom there are more than 2500 H atoms, and Fe becomes even scarcer in galaxies with lower metal content, such as the nearby Large and Small Magellanic Clouds, at respectively 1/2 and 1/5 of solar metallicity. Owing to the highly complex atomic structure of Fe, it has millions of line transitions, which makes it an extremely efficient absorber of radiation in the inner atmosphere where the mass-loss rate is set (Vink et al., 1999; Puls et al., 2000).

Up to 2005, most stellar modellers assumed that due to the overwhelming presence of carbon in WC star atmospheres, it should also be the element of carbon that drives Wolf–Rayet winds, rather than the few Fe atoms, which were basically assumed to constitute a negligible amount. This assumption also implied that WC stars in low-metallicity ($Z$) galaxies would have stellar winds equally strong as those in the Galaxy, still removing the required angular momentum. It was for this very reason that there was no satisfactory explanation for the long GRB puzzle.

Nevertheless, GRBs were found to arise in low-metallicity galaxies (e.g., Vreeswijk et al., 2004), characteristic of conditions in the early Universe. It is interesting to note that the most distant object in our Universe known today is indeed a GRB, estimated to have resulted from the collapse of a massive star only some 500 mil-lion years after the Big Bang (Tanvir et al., 2009). That GRBs occur in low-metallicity environments may imply that they were common at earlier times in the Universe when the interstellar gas was less enriched.

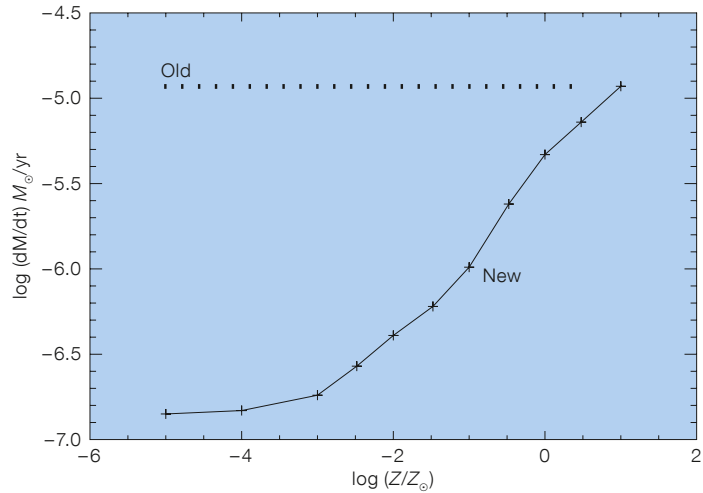

Figure 3. The mass loss versus (host galaxy) metallicity relation for late-type WC stars as found by Vink & de Koter (2005). The earlier concept of metallicity-independent rates is referred to as "old".

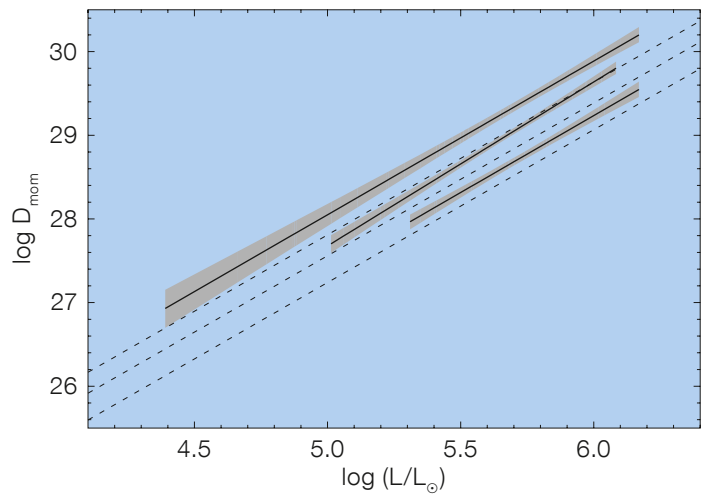

Figure 4. Comparison of the empirical wind–momentum–luminosity relations (solid lines) to theoretical predictions (dashed lines) for the Milky Way, the LMC and SMC respectively (see Mokiem et al. [2007] for details).

In 2005 Vink & de Koter performed a pilot study of Wolf–Rayet mass loss as a function of metallicity (see Figure 3) and found that, although C is the most abundant metallic element in the Wolf–Rayet atmosphere, it is the much more complex Fe element that drives the stellar wind. Thus host galaxy metallicity plays a crucial role: objects that are born with fewer Fe atoms will lose less matter by the time they reach the ends of their lives, despite their large CNO-type material content. The striking implication is that objects formed in the early Universe and in other low-metallicity environments can retain their angular momentum, maintaining their rapid spin towards collapse, enabling a GRB event (see Figure 4). More recent stellar evolution models involving very rapidly spinning stars in which the objects become almost fully mixed, such as those of Yoon & Langer (2005) and Woosley & Heger (2006), but now assuming that the mass loss depends on the original Fe metal content (rather than that of self-enriched CNO elements) indeed shows that massive stars can maintain their rapid rotation until collapse. This finding appears to resolve the collapsar puzzle for long GRBs: the outcome depends on just a very few Fe atoms! The key question is now what happens in nature.

Observational tests

The key assumptions and predictions of the collapsar model for GRBs that have to be met are that stellar winds should depend on metallicity and that lower metallicity Wolf–Rayet stars should rotate more rapidly than their higher metallicity Galactic counterparts.

The first part of the observational evidence concerns the issue of whether stellar winds are observed to depend on metallicity. For O-type stars this has been suspected for a long time and the most recent results by the FLAMES consortium on massive stars find good agreement between observed and predicted wind momenta versus stellar luminosities (see Figure 4). The situation for Wolf–Rayet stars has been more confused, with controversial results in the early 2000s, although more recently a mass-loss metallicity dependence of $\dot{M} \propto Z^{0.8}$ was suggested by Crowther (2006), in good agreement with theoretical relations (Vink & de Koter, 2005; Gräfener & Hamann, 2008).

The second observational step would be to show that WR stars at lower metallicity rotate more slowly than those in the Galaxy. For most stellar applications one would simply measure the rotation rate (actually $v\sin i$) from the width of stellar absorption lines, but in WR stars the spectra are dominated by emission lines, and it is not clear whether the line shape arises from rotation or from other dynamical effects associated with the outflow. In other words, the route of stellar spectroscopy is unsuitable for measuring rotation rates directly in WR winds. Fortunately, there is an alternative method available via measurement of the polarisation properties across emission lines, in the form of linear spectropolarimetry.

The tool of linear spectropolarimetry

Linear polarisation is a very powerful technique to deduce the presence of asymmetric structures, even in cases where the objects under consideration cannot be spatially resolved. The amount of linear polarisation is simply given by the vector sum of the Stokes parameters $Q$ and $U$. In many instances however, interstellar dust grains produce additional polarisation, and this would normally not occur at exactly the same position angle (see Figure 5). If the interstellar and the observed polarisations can be disentangled, the amount of polarisation from the source can be measured directly, and it is easy to infer whether any given object is spherically symmetric or not. One way to achieve this is by performing spectropolarimetry.

In its simplest form, the technique is based on the expectation that line photons arise over a larger volume than continuum photons. If so, then line photons undergo fewer scatterings from, for example, a flattened wind than continuum photons, and the emission-line flux becomes less polarised, resulting in a polarisation variation across the line (see Figure 7).

The high incidence (~ 60 %) of "line effects" revealed among classical Be stars in the 1970s indicated that all classical Be stars possess discs, with orientation towards the observer ($\sin i$) determining whether any given object is subject to a line effect or not. This serves to show that significantly sized samples are needed for meaningful analysis. Another criterion is high signal-to-noise, as spectropolarimetry is a "photon-hungry" technique, and the tool is thus best served by the largest telescopes, such as ESO's VLT.

VLT/FORS spectropolarimetry programme of low-metallicity WR stars

Over the last couple of years, we have utilised the tool of linear spectropolarimetry on WR stars in the Large and Small Magellanic Clouds (LMC & SMC) to test the assumptions of the collapsar model for long GRBs. This has been generously supported in Periods 78, 81 and 85.

If the WR mass-loss metallicity dependence, and the subsequent inhibition of angular momentum removal are the key to explaining the high occurrence of GRBs at low metallicity, WR stars in the Magellanic Clouds should, on average spin faster than those in the Galaxy. A linear spectropolarimetry survey of LMC WR stars (see examples in Figures 6 and 7) showed that ~ 15 % of LMC WR stars bear the sign of rapid rotation, as only two out of 13 of them show a significant amount of linear polarisation (Vink, 2007). The incidence rate is equal to that of the Galactic WR survey by Harries et al. (1998). The LMC sample, however, was necessarily biased towards the brightest objects (with $V_{\mathrm{magnitude}}$ < 12, mostly containing very late type nitrogen-rich WR stars and/or binaries), and for an unbiased assessment we have been allocated time to study a larger sample.

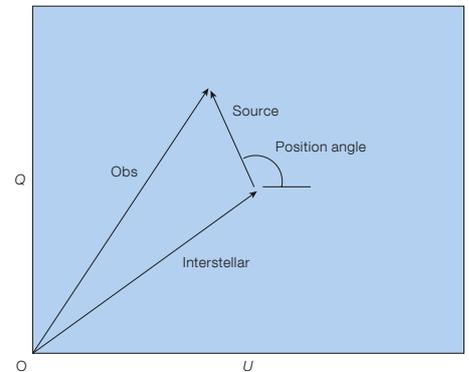

Figure 5. A schematic $QU$ diagram. The observed linear polarisation is a vector addition of the interstellar plus source polarisation. The length of the vectors represents the degree of polarisation, whilst the angle provides the position angle $\theta$.

The data obtained thus far may suggest that the metal content of the LMC is high enough for WR winds to remove the necessary angular momentum, and single star progenitors may be constrained to an upper metallicity of that of the LMC at 1/2 solar. Alternatively, rapid rotation of WR-type objects may only be achievable for objects that are the products of binary evolution. Meaningful correlations between binarity and large amounts of linear polarisation have yet to be performed, which is one of the main tasks of our ongoing VLT/FORS programme.

Future testing with the E-ELT?

Looking further ahead, with linear spectropolarimetry on a 42-metre European Extremely Large Telescope (E-ELT), one could properly probe WR wind geometries for very faint objects such as those in the very low metallicity SMC (at 1/5 solar), and beyond. The aim is to constrain GRB progenitor models in the criti-cal metallicity range between 1/5 and 1/2 solar, which is anticipated to have crucial implications for understanding the production of GRBs at low metallicity in the early Universe.

Future Extremely Large Telescope (ELT) work could also involve a complete census of all evolved massive supernova progenitors (Wolf–Rayet stars, B[e] supergiants, and Luminous Blue Variables) in both the LMC and SMC, in order to map the links between mass loss and rotation in the most massive stars.

If the spectral resolution of a potential E-ELT spectropolarimeter were ultra-high, one could also perform complementary circular spectropolarimetry to measure magnetic fields in these low-metallicity massive stars, and study the intricate interplay between mass loss, rotation and magnetic fields, in order to constrain massive star models as a function of $Z$, with unique constraints on models of GRBs and Pair–Instability SNe in the early Universe.

Massive stars play a vital role in the Universe as providers of ionising radiation, kinetic energy, and heavy elements. Their evolution towards collapse is driven by mass loss and rotation. As the favoured progenitors of long GRBs, massive stars may also be our best signpost of individual objects in the early Universe, but we do not know why certain massive stars collapse to produce GRBs while others do not. Studies of wind asymmetries in massive stars are vital for understanding massive star evolution, and thereby GRB production and related phenomena.

References

Crowther, P. A. 2006, ASPC, 353, 157
Galama, T. J. et al. 1998, Nature, 395, 670
Gräfener, G. & Hamann, W.-R. 2008, A&A, 482, 945
Harries, T. J. et al. 1998, MNRAS, 296, 1072
Langer, N. 1998, A&A, 329, 551
Meynet, G. & Maeder, A. 2003, A&A, 404, 975
Mokiem, M. R. et al. 2007, A&A, 473, 603
Puls, J. et al. 2000, A&AS, 141, 23
Tanvir, N. et al. 2009, Nature, 461, 1254
Vink, J. S. 2007, A&A, 469, 707
Vink, J. S. & de Koter, A. 2005, A&A, 442, 587
Vink, J. S. et al. 1999, A&A, 345, 109
Vreeswijk, et al. 2004, A&A, 419, 927
Woosley, S. E. & Heger, A. 2006, ApJ, 637, 914
Woosley, S. E. 1993, ApJ, 405, 273
Yoon, S.-C. & Langer, N. 2005, A&A, 443, 643

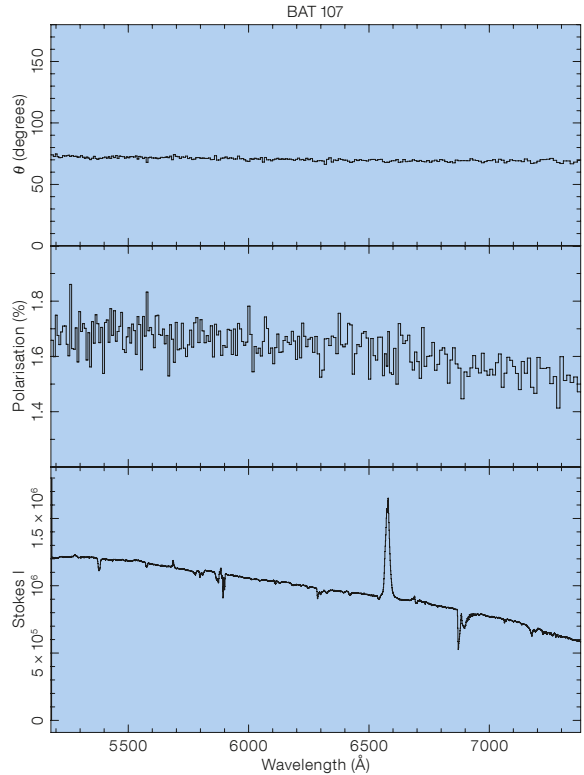

Figure 6. Linear polarisation triplot for the LMC Wolf–Rayet star BAT 107. The upper panel shows the polarisation position angle, the middle panel the degree of polarisation, and the lower panel the normal Stokes I spectroscopy. BAT 107 does not show any "line effects" (viz. polarisation signature across the lines) and thus no evidence for intrinsic polarisation, nor asphericity (Vink, 2007).

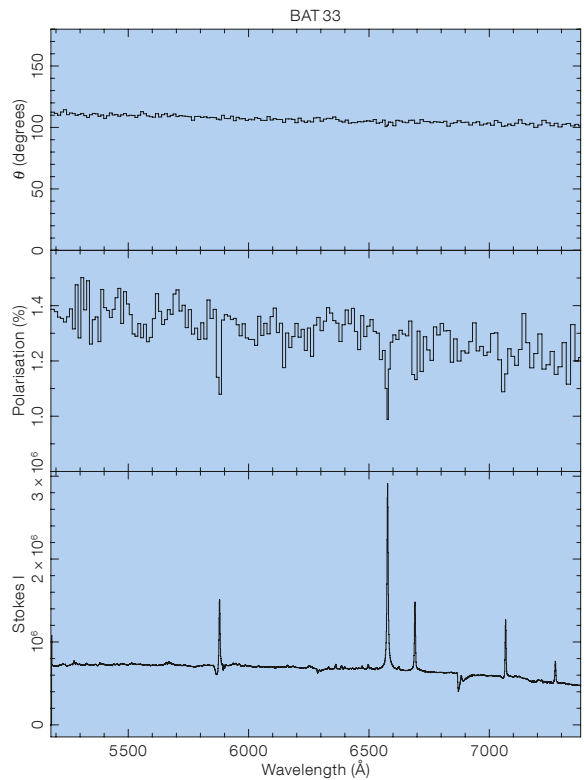

Figure 7. Linear polarisation triplot for the LMC Wolf–Rayet star BAT 33, layout as Figure 4. BAT 33 clearly shows some "line effects" and thus clear-cut evidence for intrinsic polarisation and asymmetry (Vink, 2007).